\documentclass[5p]{elsarticle}

\usepackage{multirow,color,hyperref}

\journal{New Astronomy}

\usepackage{numcompress}\biboptions{authoryear}

\begin{document}

\begin{frontmatter}

\title{Analysis of the observed and intrinsic durations of \textit{Swift}/BAT gamma-ray bursts}
\author{Mariusz Tarnopolski}
\address{Astronomical Observatory, Jagiellonian University, Orla 171, Krak\'ow, Poland}
\ead{mariusz.tarnopolski@uj.edu.pl}

\begin{abstract}
The duration distribution of 947 GRBs observed by {\it Swift}/BAT, as well as its subsample of 347 events with measured redshift, allowing to examine the durations in both the observer and rest frames, are examined. Using a maximum log-likelihood method, mixtures of two and three standard Gaussians are fitted to each sample, and the adequate model is chosen based on the value of the difference in the log-likelihoods, Akaike information criterion and Bayesian information criterion. It is found that a two-Gaussian is a better description than a three-Gaussian, and that the presumed intermediate-duration class is unlikely to be present in the {\it Swift} duration data.
\end{abstract}

\begin{keyword}
gamma-ray burst: general\sep methods: data analysis\sep methods: statistical
\end{keyword}

\end{frontmatter}

\section{Introduction}\label{intro}

Gamma-ray bursts (GRBs) were detected by military satellites \textit{Vela} in late 1960's. \citet{mazets} first pointed out hints for a bimodal distribution of $T_b$ (taken to be the time interval within which fall $80\%-90\%$ of the measured GRB's intensity) drawn for 143 events detected in the KONUS experiment. Burst and Transient Source Explorer (BATSE) onboard the Compton Gamma Ray Observatory (\textit{CGRO}) provided data that were further investigated by \citet{kouve}, and led to establishing the common classification of GRBs into short ($T_{90}<2\,{\rm s}$) and long ($T_{90}>2\,{\rm s}$), where $T_{90}$ is the time during which 90\% of the burst's fluence is accumulated, referred to as the duration of a GRB. The progenitors of long GRBs are associated with supernovae related with collapse of massive stars \citep{woosley}. Progenitors of short GRBs are thought to be NS-NS or NS-BH mergers \citep{nakar}, and no connection between short GRBs and supernovae has been proven \citep{zhang5}. It was observed that durations $T_{90}$ seem to exhibit log-normal distributions which were thereafter fitted to short and long GRBs \citep{mcbreen,koshut,kouve2,horvath02}. 

The existence of an intermediate-duration GRB class, consisting of GRBs with $T_{90}$ in the range $2-10\,{\rm s}$, was put forward \citep{horvath98,mukh} based on the analysis of BATSE 3B data. It was supported \citep{horvath02,chatto} with the use of the complete BATSE dataset. Evidence for a third log-normal component was also found in {\it Swift}/BAT data \citep{horvath08,zhang2,huja,horvath10}. Interestingly, \citet{zitouni} re-examined the BATSE current catalog as well as the {\it Swift} dataset, and found that a mixture of three Gaussians (3-G) fits the $\log T_{90}$ data from {\it Swift} better than a two-Gaussian (2-G), while in the case of BATSE statistical tests did not support the presence of a third component (hereinafter, the $\log T_{90}$ distributions are considered, and are shortly referred to as durations as well). Regarding {\it Fermi}/GBM \citep{gruber,kienlin}, a 3-G is a better fit than a 2-G,\footnote{Adding parameters to a nested  model always results in a better fit (in the sense of a lower $\chi^2$ or a higher maximum log-likelihood) due to more freedom given to the model to follow the data, i.e. due to introducing more free parameters. The important question is whether this improvement is statistically significant, and whether the model is justified\label{fn1}.} however the presence of a third group in the duration distribution was found to be unlikely \citep{Tarnopolski,Tarnopolski2}, which was based on the fact that the $\log T_{90}$ distribution is bimodal, i.e. it exhibits two local maxima \citep{Tarnopolski}, and that a mixture of two skewed components follows the data better than a standard three-Gaussian \citep{Tarnopolski2}.

The {\it Swift} data were re-examined by \citet{bromberg}, and they found that a limit of $0.8\,{\rm s}$ is more suitable for the GRBs observed by {\it Swift} than the conventional $2\,{\rm s}$ limit of \citet{kouve}. It should be stressed that \citet{bromberg} applied a different approach than \citet{kouve} and \citet{Tarnopolski3}: a functional form of the $T_{90}$ distribution different from the commonly used phenomenological log-normal distribution, coming from a physical model for the short duration collapsar distribution, and by means of exceeding a probability threshold that a GRB with a given $T_{90}$ is a non-collapsar. Interestingly, the limits for BATSE and {\it Fermi} data are consistent with the $2\,{\rm s}$ limit, and also with the results obtained by \citet{Tarnopolski3}, where based on the well-established conjecture that durations $T_{90}$ are log-normally distributed, the limit between short and long GRBs may be placed at the local minimum, which is detector-dependent. Finally, many works in which a 2-G was fitted to the $\log T_{90}$ distribution showed a significant overlap of components corresponding to short and long GRBs \citep{mcbreen,koshut,horvath02,zhang2,huja,bromberg,barnacka,Tarnopolski3,zitouni}.

The aim of this paper is to analyze the current dataset of {\it Swift}/BAT GRBs, and to test whether a greater sample of 947 events leads to conclusions other than \citet{zitouni} arrived at for a set of 757 events. Moreover, a relevant increase of GRBs with measured redshift---347 compared to 248 GRBs examined by \citet{zitouni}---provides an opportunity for a re-evaluation of the GRB properties that are, after moving to the rest frame, not affected by cosmological factors. This paper is organized in the following manner. In Section~\ref{meth} the datasets, fitting method and statistical criteria used to infer the validity of the models applied are described. Section~\ref{res} presents the results of fitting a 2-G and 3-G to the whole sample of 947 GRBs, as well as a subsample of 347 events in both the observer and rest frames. Section~\ref{disc} is devoted to discussion, and gathers concluding remarks.

\section{Methods}\label{meth}

\subsection{Dataset}\label{data}

\begin{figure}
\includegraphics[width=0.5\textwidth]{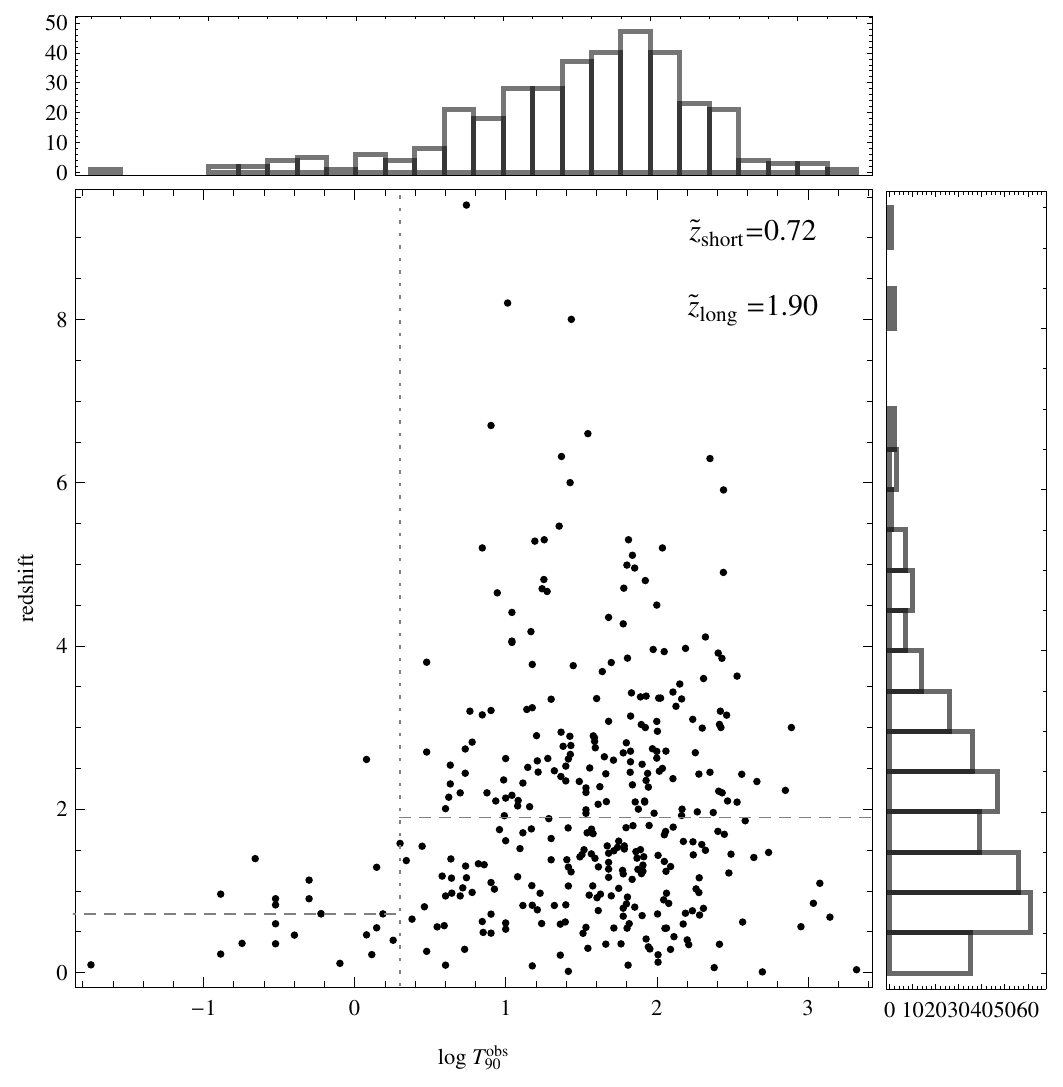}
\caption{A scatter plot of the redshift versus the observed duration of the {\it Swift} subsample. Vertical dotted line denotes the limitting value of $2\,{\rm s}$ between short and long GRBs, and the horizontal dashed lines mark the medians of the respective classes, with values written in the plot.}
\label{fig1}
\end{figure}

\begin{figure}
\includegraphics[width=0.5\textwidth]{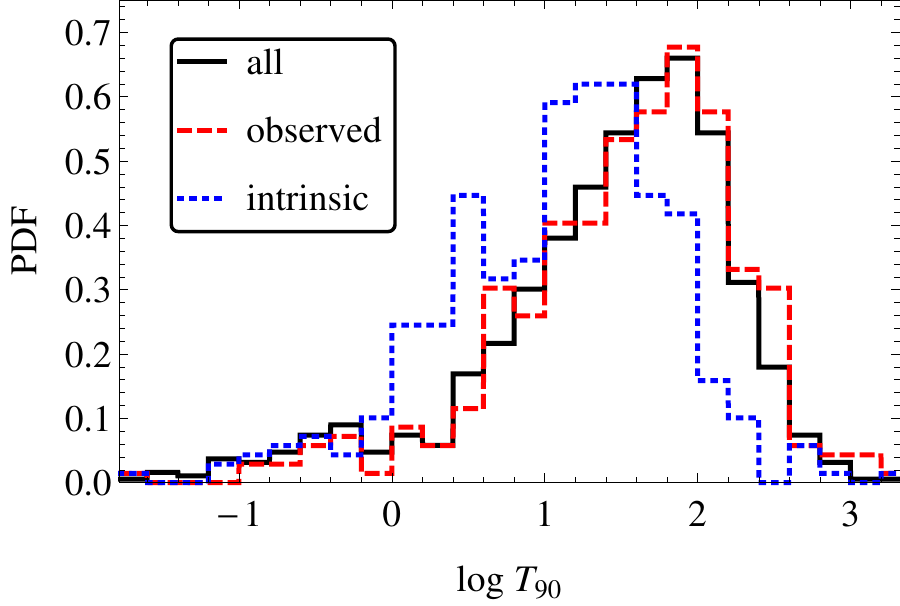}
\caption{Distributions of the examined samples: the whole sample (solid black); observed (dashed red) and intrinsic (dotted blue) durations in the subsample of GRBs with known redshift. The distributions of observed durations for both samples (all GRBs and those with measured redshift) are similar to each other.}
\label{fig2}
\end{figure}

The {\it Swift} dataset contains 947 GRBs\footnote{\url{http://swift.gsfc.nasa.gov/archive/grb\_table.html}, accessed on September 30, 2015.} with measured duration $T_{90}$, of which 9\% are short (87 events). 347 GRBs have their redshift known, and those constitute the second sample examined herein. It consists of 324 long GRBs and 23 short ones. A scatter plot of this subsample on a $\log T_{90}-z$ plane is drawn in Fig.~\ref{fig1}. The median redshift for short and long GRBs is equal to $\tilde{z}_{\rm short}=0.72$ and $\tilde{z}_{\rm long}=1.90$, respectively. The intrinsic durations are calculated according to
\begin{equation}
T^{\rm int}_{90}=\frac{T^{\rm obs}_{90}}{1+z}.
\label{eq1}
\end{equation}
Distributions of the $\log T_{90}$ for the observed and intrinsic durations are examined hereinafter, and are displayed together with the distribution of the whole sample in Fig.~\ref{fig2}.

\subsection{Fitting method}\label{fit}

Two standard fitting techniques are commonly applied: $\chi^2$ fitting \citep{voinov} and maximum likelihood (ML, \citealt{kendall}). For the first, data needs to be binned, and despite various binning rules are known (e.g. Freedman-Diaconis, Scott, Knuth etc.), they still leave place for ambiguity, as it might happen that the fit may be statistically significant on a given significance level for a number of binnings \citep{huja,koen,Tarnopolski}. The ML method is not affected by this issue and is therefore applied herein.  However, for display purposes, the binning was chosen based on the Freedman-Diaconis rule.

Having a distribution with a probability density function (PDF) given by $f=f(x;\theta)$ (possibly a mixture), where $\theta=\left\{\theta_i\right\}_{i=1}^p$ is a set of parameters, the log-likelihood function is defined as
\begin{equation}
\mathcal{L}_p(\theta)=\sum\limits_{i=1}^N\ln f(x_i;\theta),
\label{eq2}
\end{equation}
where $\left\{x_i\right\}_{i=1}^N$ are the datapoints from the sample to which a distribution is fitted. The fitting is performed by searching a set of parameters $\hat{\theta}$ for which the log-likelihood is maximized. When nested models are considered, the maximal value of the log-likelihood function $\mathcal{L}_{\rm max}\equiv\mathcal{L}_p(\hat{\theta})$ increases when the number of parameters $p$ increases.

A mixture of $k$ standard normal (Gaussian) distributions:
\begin{equation}
f_k(x) = \sum\limits_{i=1}^k \frac{A_i}{\sqrt{2\pi}\sigma_i}\exp\left(-\frac{(x-\mu_i)^2}{2\sigma_i^2}\right),
\label{eq9}
\end{equation}
is considered. It is described by $p=3k-1$ free parameters: $k$ pairs $(\mu_i,\sigma_i)$ and $k-1$ weights $A_i$, satysfying $\sum_{i=1}^k A_i=1$ due to normalization of a PDF. Therefore, $p=5$ for a 2-G, and $p=8$ for a 3-G.

\subsection{Statistical criteria}\label{crit}

If one has two fits such that $\mathcal{L}_{p_2,{\rm max}} > \mathcal{L}_{p_1,{\rm max}}$, then twice their difference, $2\Delta\mathcal{L}_{\rm max}=2(\mathcal{L}_{p_2,{\rm max}}-\mathcal{L}_{p_1,{\rm max}})$, is distributed like $\chi^2(\Delta p)$, where $\Delta p=p_2-p_1>0$ is the difference in the number of parameters \citep{kendall,horvath02}. If a $p$-value associated with the value of $\chi^2(\Delta p)$ does not exceed the significance level $\alpha$, one of the fits (with higher $\mathcal{L}_{\rm max}$) is statistically better than the other. For instance, for a 2-G and a 3-G, $\Delta p=3$, and despite that, according to Footnote~\ref{fn1}, $\mathcal{L}_{\rm max,\,3-G} > \mathcal{L}_{\rm max,\,2-G}$ holds always, twice their difference provides a decisive $p$-value.

For nested as well as non-nested models, the Akaike information criterion ($AIC$) \citep{akaike,burnham,liddle} may be applied. The $AIC$ is defined as
\begin{equation}
AIC=2p-2\mathcal{L}_{p,{\rm max}}.
\label{eq3}
\end{equation}
A preferred model is the one that minimizes $AIC$. The formulation of $AIC$ penalizes the use of an excessive number of parameters, hence discourages overfitting. It prefers models with fewer parameters, as long as the others do not provide a substantially better fit. The expression for $AIC$ consists of two competing terms: the first measuring the model complexity (number of free parameters), and the second measuring the goodness of fit (or more precisely, the lack of thereof). Among candidate models with $AIC_i$, let $AIC_{\rm min}$ denote the smallest. Then,
\begin{equation}
Pr_i=\exp\left(-\frac{\Delta_i}{2}\right),
\label{eq4}
\end{equation}
where $\Delta_i=AIC_i-AIC_{\rm min}$, can be interpreted as the relative (compared to $AIC_{\rm min}$) probability that the $i$th model minimizes the $AIC$.\footnote{Relative probabilities normalized to unity are called the Akaike weights, $w_i=\frac{\exp\left(-\Delta_i/2\right)}{\sum\limits_i \exp\left(-\Delta_i/2\right)}$. In Bayesian language, Akaike weight corresponds to the posterior probability of a model (under assumption of different prior probabilities; see \citealt{biesiada}).}

The $AIC$ is suitable when $N/p$ is large, i.e. when $N/p>40$ \citep[][see also references therein]{burnham}. When this condition is not fulfilled, a second order bias correction is introduced, resulting in a small-sample version of the $AIC$, called $AIC_c$:
\begin{equation}
AIC_c=2p-2\mathcal{L}_{p,{\rm max}}+\frac{2p(p+1)}{N-p-1}.
\label{eq5}
\end{equation}
The relative probability is computed similarly to when $AIC$ is used, i.e. Eq.~(\ref{eq4}) is valid when one takes $\Delta_i=AIC_{c,i}-AIC_{c,{\rm min}}$. Thence,
\begin{equation}
Pr_i=\exp\left(-\frac{AIC_{c,i}-AIC_{c,{\rm min}}}{2}\right).
\label{eq6}
\end{equation}
Obviously, $AIC_c$ converges to $AIC$ when $N$ is large.

It is important to note that this method allows to choose a model that is best among a given set, but does not allow to state that this model is the best among all possible ones. Hence, the probabilities computed by means of Eq.~(\ref{eq6}) are the relative (with respect to a model with $AIC_{c,{\rm min}}$) probabilities that the data is better described by a model with $AIC_{c,i}$. What is essential in assesing the goodness of a fit in the $AIC$ method is the difference, $\Delta_i=AIC_{c,i}-AIC_{c,{\rm min}}$, not the absolute value of an $AIC_{c,i}$.\footnote{The $AIC$ value contains scaling constants coming from the log-likelihood $\mathcal{L}$, and so $\Delta_i$ are free of such constants \citep{burnham}. One might consider $\Delta_i=AIC_{c,i}-AIC_{c,{\rm min}}$ a rescaling transformation that forces the best model to have $\Delta_{\rm min}:=0$.} If $\Delta_i<2$, then there is substantial support for the $i$th model, and the proposition that it is a proper description is highly probable. If $2<\Delta_i<4$, then there is strong support for the $i$th model. When $4<\Delta_i<7$, there is considerably less support, and models with $\Delta_i>10$ have essentially no support \citep{burnham}.

The Bayesian information criterion ($BIC$) was introduced by \citet{schwarz}, and is defined as
\begin{equation}
BIC=p\ln N-2\mathcal{L}_{p,{\rm max}}.
\label{eq7}
\end{equation}
As was the case in  the $AIC$ (or $AIC_c$), a preferred model is the one that minimizes $BIC$, which also penalizes the usage of an excessive number of free parameters. The most striking difference between the two is that the penalization term, $k\ln N$, is greater than the corresponding term from the $AIC$, i.e. $2p$, for $N\geq 8$. Hence, the penalization in case of the $BIC$ is much more stringent, especially for large samples.

The probability in favor of the $i$th model, relative to a model with $BIC_{\rm min}$, is defined in the same manner as it was for $AIC$:
\begin{equation}
Pr_i=\exp\left(-\frac{\Delta_i}{2}\right),
\label{eq8}
\end{equation}
where $\Delta_i=BIC_i-BIC_{\rm min}$ in this case, and the support for the $i$th model (or evidence against it) also depends on the differences: if $\Delta_i<2$, then there is substantial support for the $i$th model (or the evidence against it is worth only a bare mention). When $2<\Delta_i<6$, then there is positive evidence against the $i$th model. If $6<\Delta_i<10$, the evidence is strong, and models with $\Delta_i>10$ yield a very strong evidence against the $i$th model (essentially no support, \citealt{kass}).

Despite apparent similarities between the $AIC$ and $BIC$, they answer different questions, as they are derived based on different assumptions. $AIC$ tries to select a model that most adequately describes reality (in the form of the data under examination). This means that in fact the model being a real description of the data is never considered. On the contrary, $BIC$ tries to find the true model among the set of candidates. Because $BIC$ is more stringent, it has a tendency to underfit, while $AIC$, as a more liberal method, is inclined towards overfitting. This leads sometimes to pointing different models by the two criteria, which happens rarely, but is due to the fact that they try to satisfy different conditions.

\section{Results}\label{res}

\subsection{All \textit{Swift} GRBs}\label{res1}

\begin{figure}
\includegraphics[width=0.5\textwidth]{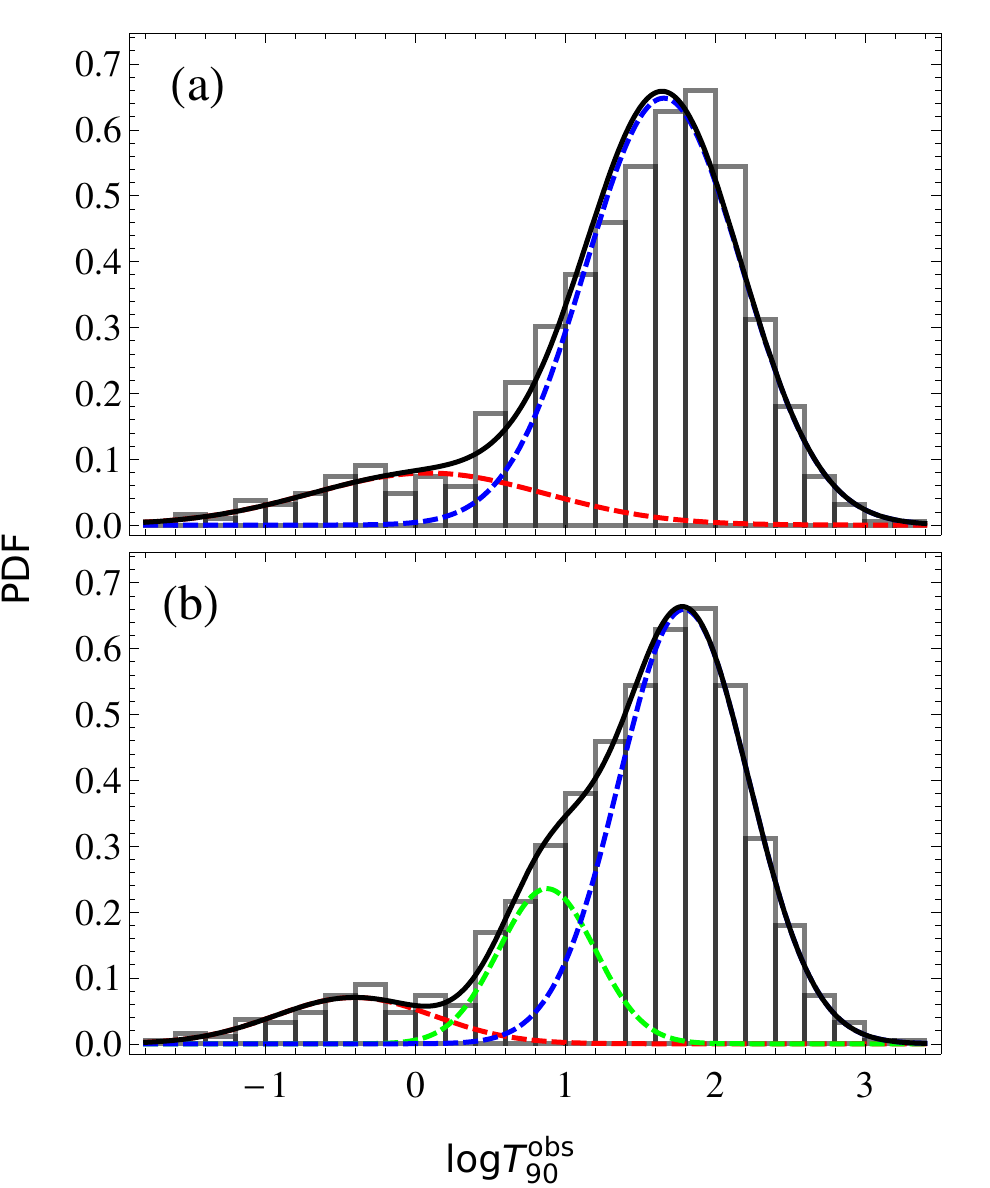}
\caption{Distributions fitted to $\log T^{\rm obs}_{90}$ of all {\it Swift} GRBs. Color dashed curves are the components of the (black solid) mixture distribution. The panels show mixtures of (a) two and (b) three standard Gaussians.}
\label{fig3}
\end{figure}

Using the ML method from Sect.~\ref{fit}, the fitting of a 2-G and 3-G to the duration distribution of 947 GRBs is performed. The results, in graphical form, are shown in Fig.~\ref{fig3}. The two-component mixture is unimodal, but with a prominent tail on the left side. The three-component fit is bimodal, with a clear shoulder on the left side of the peak related to long GRBs. The local minimum is placed at $T_{90}=1.07\,{\rm s}$. It is interesting to note that this value is consistent with the short-long GRB limit from \citet{bromberg}. The overall shape of the curve is in agreement with previous results \citep{horvath08,zhang2,Tarnopolski2,zitouni}.

The parameters of the fits are gathered in Table~\ref{tbl1}. Twice the difference in $\mathcal{L}_{p,{\rm max}}$ is equal to 22.336, what corresponds to a $p$-value of $6\times 10^{-5}$, indicating that a 3-G is a highly significant improvement over a 2-G. This is confirmed with the $AIC_c$ approach, as their difference is equal to 16.246, what gives a probability of $3\times 10^{-4}$ that the 2-G might in fact be a better description than a 3-G. However, the results of the $BIC$ give a much lower significance---the difference is only 1.776, corresponding to a probability of 0.41 in favor of the 2-G. Nevertheless, all criteria pointed at a 3-G as a better model among the two under consideration.

\begin{table}
\small
\caption{Parameters of the fits for the observed durations of 947 {\it Swift} GRBs. The values in favor of a respective model are marked in bold.}
\label{tbl1}
\centering
\begin{tabular}{@{}ccccccc@{}}
\hline\hline
$i$ & $\mu_i$ & $\sigma_i$ & $A_i$ & $\mathcal{L}_{\rm max}$ & $AIC_c$ & $BIC$\\
  \hline
1 & 0.083 & 0.781 & 0.155 & \multirow{2}{*}{$-1035.351$} & \multirow{2}{*}{2080.765} & \multirow{2}{*}{2104.968}\\
2 & 1.657 & 0.521 & 0.845 & 							 & 							 & 							\\
  \hline
1 & $-0.407$ & 0.529 & 0.093 &			 &		    & \\
2 &  0.878   & 0.322 & 0.190 & $-$\textbf{1024.183} & \textbf{2064.519} & \textbf{2103.192} \\
3 &  1.793   & 0.434 & 0.717 & 			 &		    & \\
  \hline
\end{tabular}
\end{table}

\subsection{Subsample of \textit{Swift} GRBs with measured redshift}\label{res2}

\subsubsection{Observed durations}\label{res21}

\begin{figure}
\includegraphics[width=0.5\textwidth]{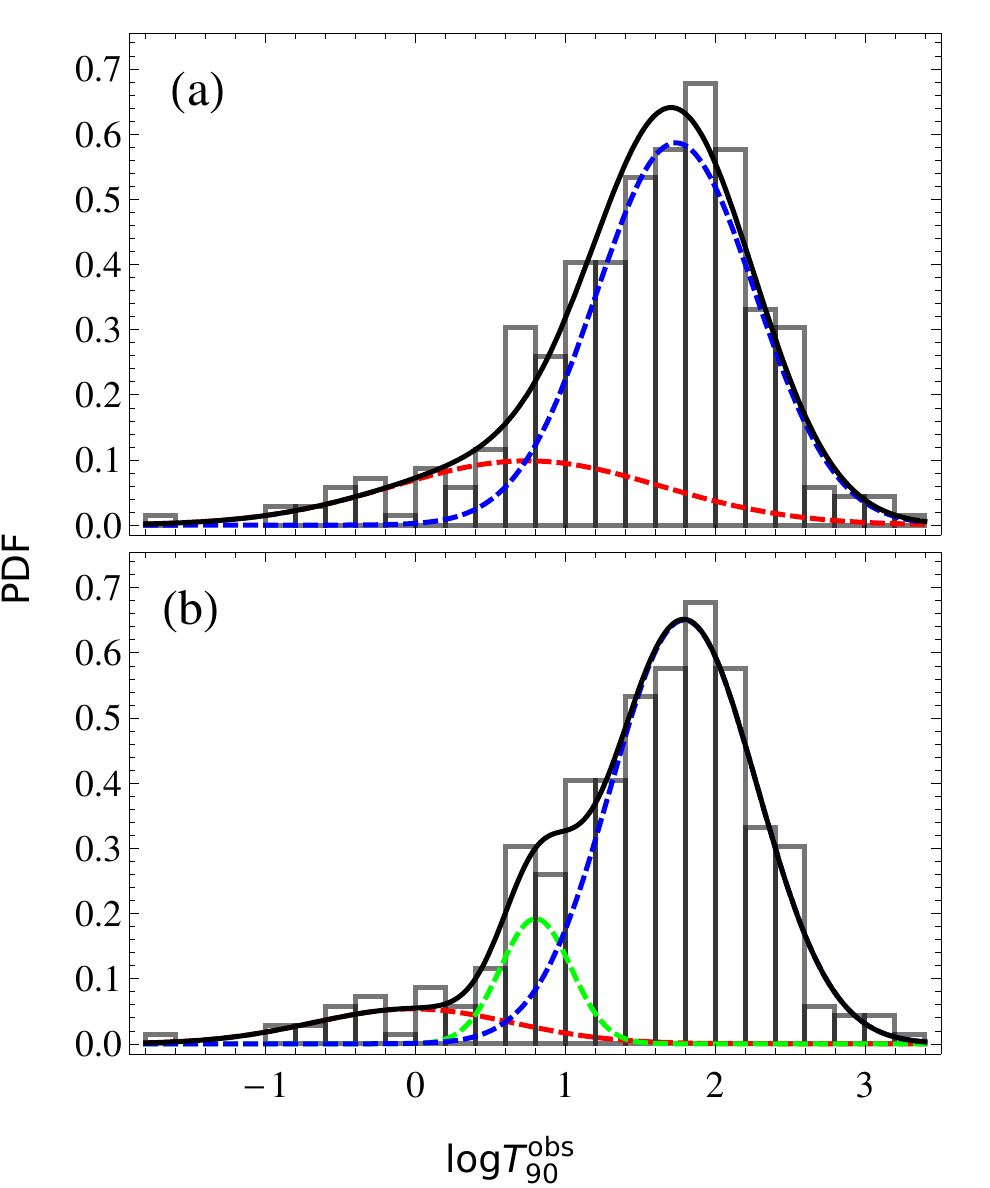}
\caption{The same as Fig.~\ref{fig3}, but for a subsample of GRBs with measured redshift.}
\label{fig4}
\end{figure}

The observed durations of the redshift-equipped GRB subsample are examined in the same way as in the previous Section~\ref{res1}. The fitted curves, displayed in Fig.~\ref{fig4}, resemble the ones obtained for the complete GRB sample. The fitted parameters, gathered in Table~\ref{tbl2}, are in good agreement, too. The difference in $\mathcal{L}_{p,{\rm max}}$ multiplied by two, being equal to 6.522, corresponds to a relatively high probability of 0.09. The difference in $AIC_c$, equal to 0.270, is negligible, and hence based on it one can not rule out any of the fits---the relative probability is 0.87. On the other hand, the lower $BIC$ was achieved by a 2-G, and the difference is a prominent 11.027, which gives a probability of $4\times 10^{-3}$. Hence, the overall evidence against a 3-G is very strong---it follows that based on $BIC$, a 2-G is a better model.
\begin{table}
\small
\caption{Parameters of the fits for the observed durations of 347 {\it Swift} GRBs with measured redshift.}
\label{tbl2}
\centering
\begin{tabular}{@{}ccccccc@{}}
\hline\hline
$i$ & $\mu_i$ & $\sigma_i$ & $A_i$ & $\mathcal{L}_{\rm max}$ & $AIC_c$ & $BIC$\\
  \hline
1 & 0.745 & 0.894 & 0.221 & \multirow{2}{*}{$-372.281$} & \multirow{2}{*}{754.737} & \multirow{2}{*}{\textbf{773.808}}\\
2 & 1.739 & 0.530 & 0.779 & & & \\
  \hline
1 & $-0.019$ & 0.664 & 0.089 &					 &		    & \\
2 &  0.803   & 0.236 & 0.114 & $-$\textbf{369.020} & \textbf{754.467}  & 784.835 \\
3 &  1.793   & 0.434 & 0.797 & 					 &		    & \\
  \hline
\end{tabular}
\end{table}

\subsubsection{Intrinsic durations}\label{res22}

\begin{figure}
\includegraphics[width=0.5\textwidth]{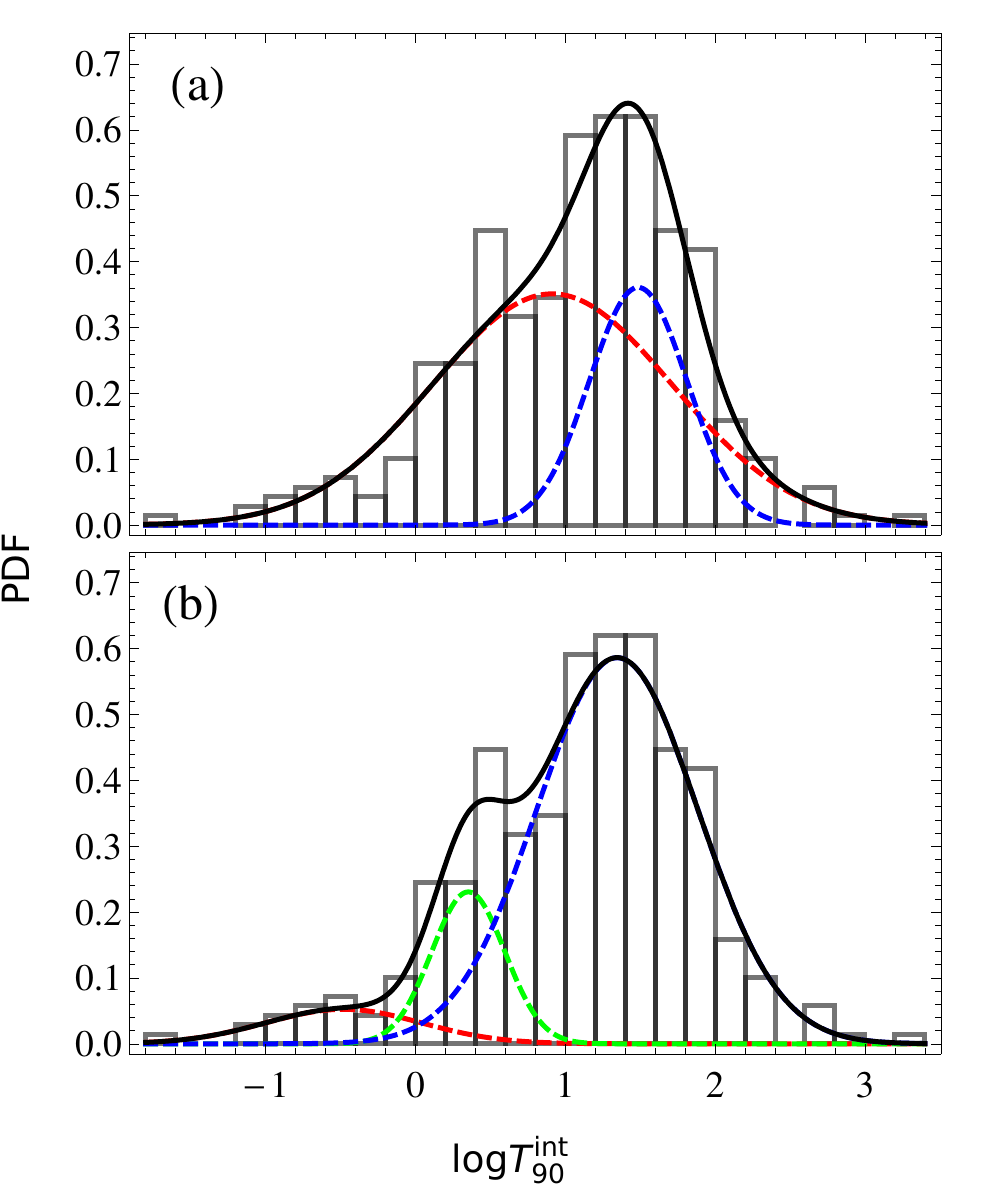}
\caption{The same as Fig.~\ref{fig4}, but in the rest frame.}
\label{fig5}
\end{figure}

In the case of the intrinsic durations, the fits displayed in Fig.~\ref{fig5} reveal a systematic shift, compared to the distribution of $T^{\rm obs}_{90}$, towards shorter durations. While a 2-G is again unimodal and skewed leftwards, the 3-G shows a peak at $T_{90}=1.42\,{\rm s}$, between the regions of short and long GRBs. The parameters of the fits are gathered in Table~\ref{tbl3}.

The doubled difference of $\mathcal{L}_{p,{\rm max}}$ is equal to 1.504, what implies a high relative probability of 0.68. The difference in $AIC_c$ is 4.747, what hints toward a 2-G with a relative probability of 0.09 that a 3-G is in fact a better model. In this case, the $BIC$ yield a difference of 16.045, which gives a strong support in favor of a 2-G. The relative probability that the 3-G might be the more appropriate description of the data, is only $3\times 10^{-4}$. Overall, the $AIC_c$ and $BIC$ are in agreement, and without taking into account the doubled difference of $\mathcal{L}_{p,{\rm max}}$, because it provides no evidence in favor of any model, it turns out that a 2-G is a better model than a 3-G.
\begin{table}
\small
\caption{Parameters of the fits for the intrinsic durations of 347 {\it Swift} GRBs.}
\label{tbl3}
\centering
\begin{tabular}{@{}ccccccc@{}}
\hline\hline
$i$ & $\mu_i$ & $\sigma_i$ & $A_i$ & $\mathcal{L}_{\rm max}$ & $AIC_c$ & $BIC$\\
  \hline
1 & 0.913 & 0.802 & 0.705 & \multirow{2}{*}{$-379.644$} & \multirow{2}{*}{\textbf{769.463}} & \multirow{2}{*}{\textbf{788.534}}\\
2 & 1.487 & 0.326 & 0.295 & & & \\
  \hline
1 & $-0.480$ & 0.521 & 0.068 &			  		 &		    & \\
2 &  0.353   & 0.244 & 0.141 & $-$\textbf{378.892} & 774.210  & 804.579 \\
3 &  1.347   & 0.538 & 0.791 & 			   		 &		    & \\
  \hline
\end{tabular}
\end{table}

\section{Discussion and conclusions}\label{disc}

The duration distribution of 947 GRBs observed by {\it Swift}, and a subsample consisting of 347 events with measured redshift, were investigated. The redshifts allowed to examine the intrinsic durations, i.e. in the rest frame, as well. Mixtures of two and three standard Gaussians were fitted. For each sample, the best fit was chosen based on the value of the difference in the log-likelihoods doubled, Akaike information criterion and Bayesian information criterion. The main conclusions are as follows:
\begin{enumerate}
\item All three criteria point at a 3-G as an adequate description of the $\log T_{90}$ distribution of all 947 {\it Swift} GRBs. The fit is bimodal, what is in good agreement with the two well established populations (mergers for short, and collapsars for long GRBs). This might suggest that the commonly applied log-normal distribution is not a good model for the observed duration distribution.
\item For a subsample of 347 GRBs with measured redshift, the analyses of the observed and intrinsic durations yielded results being in quite good agreement. While only the $BIC$ hints at a unimodal 2-G in the case of $\log T^{\rm obs}_{90}$, the other two criteria did not yield support for any of the models strong enough to infer their plausibility, but with a low ratio of short to long GRBs ($<1:14$) in this {\it Swift} subsample, combined with the well known overlap of durations of short and long GRBs and the relative smallness of the subsample, this is not that surprising.
\item The intrinsic durations, $\log T^{\rm int}_{90}$, are best described by a unimodal 2-G, too. Both $AIC_c$ and $BIC$ yield strong support in favor of a two-component Gaussian, while the criterion based on log-likelihoods did not provide a conclusive outcome. Hence, in all three samples the presence of two populations was found to be likely, and the evidence against a three-component description is strong.
\item In case of $T^{\rm obs}_{90}$ in the redshift-equipped subsample of 347 GRBs, the $AIC_c$ and $BIC$ lead to different conclusions: the $AIC_c$ pointed at a 3-G (with very weak evidence against a 2-G though), and $BIC$ yielded a very strong support for the 2-G. While this seems to be contradictive, it needs to be interpreted in the context of each criterion: according to $AIC_c$, a \mbox{3-G} is a slightly better choice in describing the data, and because the difference in the models is negligible, the $BIC$ strongly favors the simpler one, i.e. the one with fewer parameters. This illustrates that any statistical criterion can not be a stand-alone determinant in inferring the underlying model, but {\it (i)} needs to be interpreted in the light of its conditions, and {\it (ii)} it is useful to apply different tools and analyze their output in relation  to each other.
\end{enumerate}

\section*{References}

\bibliography{mybibfile}

\end{document}